\documentstyle[12pt]{article}
\begin{document}
%%%%%%%%%%%%%%%%%%%%%%%%%%%%%%%%%%%%%%%%%%%%%%%%%
%%%%%%%%%%%%%%%%%%%%%%%%%%%%%%%%%%%%%%%%%%%%%%%%%%
%%%%%%%%%%%%%%%%%%%%%%%%%%%%%%%%%%%%%%%%%%%%%%%%%%
\vspace*{2cm}
\begin{center}
 \Huge\bf
Ghost spinors, shadow electrons and the Deutsch Multiverse
\vspace*{0.25in}

\large

Elena V. Palesheva
\vspace*{0.15in}

\normalsize

Department of Mathematics, Omsk State University \\
644077 Omsk-77 RUSSIA
\\
\vspace*{0.5cm}
E-mail: m82palesheva@math.omsu.omskreg.ru  \\
\vspace*{0.5cm}
July 28, 2001\\
\vspace{.5in}
ABSTRACT
\end{center}

In this article a new solution of the Einstein-Dirac's equations
is presented.
There are ghost spinors, i.e. the stress-energy tensor is equal to
zero and the
current of these fields is non-zero vector. Last the ghost neutrino
was found.
These ghost spinors and shadow
particles of Deutsch are identified. And in result the ghost spinors have a
physical interpretation and solutions of the
field equations for shadow electrons as another shadow particles  are found.

\newpage

\setcounter{page}{1}

%%%%%%%%%%%%%%%%%%%%%%%%%%%%%%%%%%%%%%%%%%%%%%%%%%%%%%%%%%%%%%%%%%%%

\section*{Introduction}

If in General theory of relativity the right parts of the Einstein's
equations
without cosmological constant are equal to zero then one speaks about empty
space, or the space is emty if matter is  absence. But appears a question:
if a gravitational field is generated by the matter whence it appears
when matter disappears? This answer is simple: this is because required
to distinguish
such two at first thought interchangeables each other concepts as
substance and matter.
Matter generates given structure of World.  Herewith the matter is
not obliged to
have any energy that it  occurs when the stress-energy tensor equal to zero.
 Presence of substance
 is on the contrary characterized by the non-zero stress-energy
 tensor. The actual example of matter, which in our world shows
the object with zero
 energy and momentum, i.e. in introduced terminology it is not a substance,
 serves the neutrino's ghost. About
 their existance we can be to speak in light of received our results
and results of
 other authors  \cite{2,3,4,5}.

The world which surround us holds the ensemble of riddles, and we all time
 want to understand the nature of space, in which we are living.
 So and David Deutsch \cite{1} makes an attempt logically
to explain the phenomenon of an interference of quantum particles and
comes to a conclusion about existence of the parallel worlds, in all set
representing Multiverse \cite{1}.
More precisely speaking, the assumption of presence in {\it our}
spacetime of the shadow
photons, which identified by him with photons of {\it other} universe,
one results him in
the completed  ground of observed interferentional picture.
 But generally in this case the shadow particles have property of
objects with zero
 stress-energy tensor -- it is directly follows from the put
experiences -- and consequently
 their existence should be physically is proved. The results which
are discribed in this
 article,  namelly the parallel between quantum particles ghost and
 corresponding shadow particles, give such ground. Some
 times ago the corresponding to the ghost neutrinos solutions of
the Einstein-Dirac's
 equations  in cases plane-symmetric spacetimes \cite{2,5},
cyllindrically-symmetric
  spacetimes \cite{3},  and so in case of wave gravitational
field \cite{4} were found.
  Herewith spacetime is curved.
 In \cite{4} additionally it is showed the existence of the ghost neutrinos
 in a flat spacetime. The spacetime in this article also is flat.

\section{The description of spasetime geometry and corresponding spinor
fields}

We consider the Einstein-Dirac's equations
\begin{equation}%\label{1}
\left\{\begin{array}{l}
{\displaystyle R_{ik}-\frac{1}{2} g_{ik}R=\kappa T_{ik}}
\\
{\displaystyle  i\hbar {\gamma}^k\hspace*{-1mm}\left(
\frac{\partial\psi}{\partial x^{\scriptscriptstyle k}}-{\Gamma}_k\psi
\right)-mc\psi =0},
\end{array}\right.
\end{equation}
where
$$
T_{ik}=\frac{i\hbar c}{4}\left\{{\psi}^*{\gamma}^{(0)}{\gamma}_i
\left(\frac{\partial\psi}{\partial
x^{\scriptscriptstyle k}}-{\Gamma}_k\psi\right)-\left(
\frac{\partial {\psi}^*}{\partial x^{
\scriptscriptstyle
k}}{\gamma}^{(0)}+{\psi}^*{\gamma}^{(0)}{\Gamma}_k
\right){\gamma}_i\psi+\right.
$$
\begin{equation}\label{4}
\left.+{\psi}
^*{\gamma}^{(0)}{\gamma}_k\left(\frac{\partial\psi}
{\partial x^{\scriptscriptstyle i}}-{\Gamma}_i
\psi\right)-\left(\frac{\partial {\psi}^*}
{\partial x^{\scriptscriptstyle i}}{\gamma}^{(0)}+{
\psi}^*{\gamma}^{(0)}{\Gamma}_i\right){\gamma}_k\psi\right\}.
\end{equation}
Here $\psi$ is a bispinor, simbol ${}^*$ means the Hermite conjugation.

Spacetime geometry are discribed by flat metric
\begin{equation}\label{1}
 ds^2={dx^{\scriptscriptstyle 0}}^2+2e^{x^0}dx^{\scriptscriptstyle 0}
dx^{\scriptscriptstyle
 3}-{dx^{\scriptscriptstyle 1}}^2-{dx^{\scriptscriptstyle 2}}^2.
\end{equation}
So Riemann tensor $R^i_{\, klm}$ is zero and the left part of the Einstein's
equations also equal to zero. Owing to the above we receive zero spinor
matter stress-energy tensor $T_{ik}$.

In our formulas
$$
{\Gamma}_k=\frac{1}{4}g_{ml}\left(\frac{\partial{\lambda}^{(s)}_r}
{\partial x^{\scriptscriptstyle
k}}\,{\lambda}^l_{(s)}-{\Gamma}^l_{rk}\right)s^{mr},
$$
$$
s^{mr}=\frac{1}{2}\left({\gamma}^m{\gamma}^r-{\gamma}^r
{\gamma}^m\right),\quad
{\gamma}^k\equiv{\lambda}^k_{(i)}{\gamma}^{(i)},
$$
where ${\lambda}^k_{(i)}$ -- i-vector of tetrade, ${\gamma}^{(i)}$ --
matrixes of Dirac, for which we have the next  presentation with matrixes of
Pauly
$$
{\gamma}^{(0)}=\left[\begin{array}{cc}I&0\\ 0&-I\end{array}
\right],\quad{\gamma}^{(\alpha)}=
\left[\begin{array}{cc}0&{\sigma}_{\alpha}\\
-{\sigma}_{\alpha}&0\end{array}\right],
$$
$$
{\sigma}_1=\left[\begin{array}{cc}0&1\\ 1&0\end{array}\right],
{\sigma}_2=\left[\begin{array}{cc}0&-i\\ i&0\end{array}\right],
{\sigma}_3=\left[\begin{array}{cc}1&0\\ 0&-1\end{array}\right],
I=\left[\begin{array}{cc}1&0\\ 0&1\end{array}\right].
$$

Metric tensor of spacetime can be expressed by
vector's tetrade in the following form \cite[c.373]{6}:
$$
ds^2=\eta_{ab}\left({\lambda}^{(a)}_idx^{\scriptscriptstyle i}
\right)\left({\lambda}^{(b)}_kdx^{
\scriptscriptstyle k}\right).
$$
In this case we have
$$
{\eta}_{ab}=\left[\begin{array}{cccc}1&0&0&0\\ 0&-1&0&0\\ 0&0&-1&0\\
0&0&0&-1\end{array}\right].
$$
Then for gravitational field (\ref{1})
$$
{\lambda}^{(0)}_i=(1,0,0,e^{x^0}),\hspace*{0.3cm}
{\lambda}^{i}_{(0)}=(1,0,0,0),
$$
$$
{\lambda}^{(1)}_i=(0,1,0,0),\hspace*{0.6cm}{\lambda}^{i}_{(1)}=(0,1,0,0),
$$
$$
{\lambda}^{(2)}_i=(0,0,1,0),\hspace*{0.6cm} {\lambda}^{i}_{(2)}=(0,0,1,0),
$$
$$
\hspace*{1.0cm}{\lambda}^{(3)}_i=(0,0,0,e^{x^0}),\hspace*{0.3cm}
{\lambda}^{i}_{(3)}=(-1,0,0,e^{-x^0}).
$$
Then ${\Gamma}_1={\Gamma}_2={\Gamma}_3=0$ and
$$
{\Gamma}_0=\frac{1}{2}\left[\begin{array}{cc}0&{\sigma}_3\\
{\sigma}_3&0\end{array}\right].
$$
\section{The Ghostes}
\subsection{The ghost neutrinos}
In this section we will consider a neutrino, i.e. in the Dirac
equation (1) we must
take $m=0$.
Herewith let us expect, that
$$
\frac{\partial\psi}{\partial x^{\scriptscriptstyle 1}}=
\frac{\partial\psi}{\partial x^{
\scriptscriptstyle 2}}=\frac{\partial\psi}{\partial x^{\scriptscriptstyle
3}}=0,\quad \frac{\partial\psi}{\partial x^{\scriptscriptstyle 0}}=
\alpha\psi,
$$
where $\alpha$ is a real constant. For example, the bispinor
\begin{equation}\label{5}
  \psi=\left[\begin{array}{c}u_0\\ u_1\\ u_2\\ u_3\end{array}
\right]e^{\alpha x^0+\beta}
\end{equation}
satisfies this conditions, where $u_i,\beta$ are  complex constants.
Then the Dirac
equation (1) in our case equivalent to
$$
\left[\begin{array}{cc}I&-{\sigma}_3\\
{\sigma}_3&-I\end{array}\right]\psi=0.
$$
Hereinafter we have
\begin{equation}\label{6}
u_0=u_2,\quad u_1=-u_3.
\end{equation}
For stress-energy tensor (\ref{4}) by considering the restrictions
on spinor field
$\psi$ we get
$$
\begin{array}{l}
T_{00}={\displaystyle -\frac{i\hbar c}{\mathstrut 2}
\alpha{\psi}^*{\gamma}^{(0)}\left\{{\gamma}_0{\Gamma}_0+
{\Gamma}_0{\gamma}_0\right\}\psi }\\
T_{01}={\displaystyle -\frac{\mathstrut i\hbar c}{\mathstrut 4}
\alpha{\psi}^*{\gamma}^{(0)}\left\{{\gamma}_1{\Gamma}_0+{\Gamma}_0{
\gamma}_1\right\}\psi }\\
T_{02}={\displaystyle -\frac{\mathstrut i\hbar c}{4\mathstrut }
\alpha{\psi}^*{\gamma}^{(0)}\left\{{\gamma}_2{\Gamma}_0+{\Gamma}_0{
\gamma}_2\right\}\psi }\\
T_{03}={\displaystyle -\frac{i\hbar c\mathstrut }{\mathstrut 4}
\alpha{\psi}^*{\gamma}^{(0)}\left\{{\gamma}_3{\Gamma}_0+{\Gamma}_0{
\gamma}_3\right\}\psi }\\
T_{11}=T_{12}=T_{13}=T_{22}=T_{23}=T_{33}=0.
\end{array}
$$
And after some transformations
\begin{equation}\label{7}
\begin{array}{l}
T_{00}=T_{03}=0\\
T_{01}={\displaystyle -\frac{i\hbar c}{4}\alpha}\left(\bar{u_0},
\bar{u_1},-\bar{u_2},-\bar{u_3}\right)\left[
\begin{array}{cc}-i{\sigma}_2&0\\ 0&i{\sigma}_2\end{array}\right]\psi \\
\mathstrut \\
T_{02}={\displaystyle -\frac{i\hbar c}{4}\alpha}\left(\bar{u_0},
\bar{u_1},-\bar{u_2},-\bar{u_3}\right)\left[
\begin{array}{cc}-i{\sigma}_1&0\\ 0&i{\sigma}_1\end{array}\right]\psi.
\end{array}
\end{equation}

In result by using (\ref{6}) we insert (\ref{5}) in (\ref{7}) and
then $T_{01}=T_{02}=0$.

And finally we receive that $T_{ik}\equiv0$, i.e. we find a solution
of the Einstein-Dirac's
equation corresponding to ghost neutrinos as the current which as
known calculated by
formula:
\begin{equation}\label{8}
j^{(k)}={\lambda}^{(k)}_i{\psi}^*{\gamma}^{(0)}{\gamma}^i\psi,
\end{equation}
is non-zero:
$$
j^{(k)}=\left(2({u_0}^2+{u_1}^2)e^{2\alpha x^0+2\beta},0,0,2({u_0}^2+{u_1}^2)
e^{2\alpha x^0+2\beta}\right).
$$

\subsection{The ghost spinors}
As ${\Gamma}_1={\Gamma}_2={\Gamma}_3=0$ then Dirac's equation
takes the following form
\begin{equation}\label{9}
i\hbar \left({\gamma}^k\frac{\partial\psi}
{\partial x^{\scriptscriptstyle k}}-{\gamma}^0
{\Gamma}_0\psi\right)-mc\psi =0,
\end{equation}
After some transformation we get
$$
\left[\begin{array}{cc}I&-{\sigma}_3\\
{\sigma}_3&-I\end{array}\right]\frac{\partial\psi}
{\partial x^{\scriptscriptstyle
0}}+\left[\begin{array}{cc}0&{\sigma}_1\\ -{\sigma}_1&0\end{array}
\right]\frac{\partial\psi}
{\partial x^{\scriptscriptstyle 1}}+\left[\begin{array}
{cc}0&{\sigma}_2\\ -{\sigma}_2&0
\end{array}\right]\frac{\partial\psi}
{\partial x^{\scriptscriptstyle 2}}+e^{-x^0}\left[
\begin{array}{cc}0&{\sigma}_3\\ -{\sigma}_3&0\end{array}\right]
\frac{\partial\psi}{\partial
x^{\scriptscriptstyle 3}}-
$$
$$
-\frac{1}{2}\left[\begin{array}{cc}-I&{\sigma}_3\\
-{\sigma}_3&I\end{array}\right]\psi=-i\frac{mc}{\hbar}\psi.
$$
Then after non-difficult calculations we have
$$
\left\{\begin{array}{l}
{\displaystyle u_{3,1}+u_{1,1}-i(u_{3,2}+u_{1,2})+e^{-x^0}
(u_{2,3}+u_{1,3})=-i\frac{mc}{\mathstrut
\hbar}(u_0-u_2)}\\
{\displaystyle u_{0,0}-u_{2,0}+iu_{1,2}-u_{1,1}-e^{-x^0}u_{1,3}+
\frac12u_0-\frac12u_2=-i\frac{
\mathstrut mc}{\mathstrut \hbar}u_2}\\
{\displaystyle u_{2,1}-u_{0,1}+i(u_{2,2}-u_{0,2})-e^{-x^0}
(u_{3,3}-u_{0,3})=-i\frac{
\mathstrut mc}{\mathstrut \hbar}(u_1+u_3)}\\
{\displaystyle -u_{1,0}-u_{3,0}-iu_{0,2}-u_{0,1}+e^{-x^0}u_{0,3}
-\frac12u_1-\frac12u_3=-i\frac{
\mathstrut mc}{\hbar}u_3}\\
\end{array}\right.
$$
where $u_{i,k}$ is a partial derivation with respect to $x^k$.
Now let us assume that $u_0=u_2$, $u_1=-u_3$. Then we have only following
restrictions
$$
\left\{\begin{array}{l}
{(u_2+u_1)}_{,3}=0\\
{\displaystyle -u_{1,1}+iu_{1,2}-e^{-x^0}u_{1,3}=-i\frac{mc}{\hbar}u_2}\\
{(u_3-u_0)}_{,3}=0\\
{\displaystyle -u_{0,1}-iu_{0,2}+e^{-x^0}u_{0,3}=-i\frac{mc}{\hbar}u_3}
\end{array}\right.
$$
Besides this let us take the following condition: $u_0=-u_1=u_2=u_3=u$. And
we find
$$
\left\{\begin{array}{l}
{\displaystyle \frac{\partial u}{\partial
\mathstrut x^{\scriptscriptstyle 1}}=0}\\
{\displaystyle -i\frac{\mathstrut \partial u}
{\partial x^{\scriptscriptstyle 2}}+e^{-x^0}\frac{\partial u}{\partial x^{
\scriptscriptstyle 3}}=-i\frac{mc}{\hbar}u}
\end{array}\right.
$$
Let $\partial u/\partial x^{\scriptscriptstyle 3}=0$. Then
$$
{\displaystyle \frac{\partial u}{\partial x^{\scriptscriptstyle
2}}=\frac{mc}{\hbar}u},
$$
and $\partial u/\partial x^{\scriptscriptstyle 0}$ is free. Herewith
$$
u=\exp\left({\frac{mc}{\hbar}x^{\scriptscriptstyle 2}+
\alpha (x^{\scriptscriptstyle
0})}\right),
$$
or in more general form
\begin{equation}\label{10}
\psi=\left[\begin{array}{r}1\\ -1\\ 1\\ 1
\end{array}\right]e^{\frac{mc}{\hbar}x^
{\scriptscriptstyle 2}+\alpha (x^{\scriptscriptstyle 0})}.
\end{equation}

The (\ref{10}) is ghost iff  $T_{ik}\equiv 0$. Take  $Im\ [\alpha
(x^{\scriptscriptstyle 0})]=0$. Then $T_{ik}\equiv 0$.
Hence we have a ghost spinor. If $m=0$ then it is ghost neutrino, which is
addition to ghost in previous section.

By using (\ref{8}) we get that density of current is non-zero:
$$
j^{(k)}=\left(4e^{2\frac{mc}{\hbar}x^{\scriptscriptstyle 2}+2\alpha (x^{\scriptscriptstyle 0})},
0,0,4e^{2\frac{mc}{\hbar}x^{\scriptscriptstyle 2}+2\alpha
(x^{\scriptscriptstyle 0})}\right),
$$
i.e. in spacetime (\ref{1}) there exist the flowes with zero
energy and momentum.

\section{Parallel Universes}

In \cite{1} the some experiments with light interferention ares discribed.
The main point of the Deutsch explanation of these experiments
is suggestion about existance of shadow
photons, whence Deutsch come to the conclusion about partition of the
multiverse on set of parallel worlds. The logically building sequence of
discourses and conclusions about presence of set of
parallel universes is  clearly brought. Though there are  some problems
with interaction of worlds between itself. In particular very difficult
to agree
with conclusion about interaction of particles
with their own shadow particles only. Actually this is only desired
suggestion, which from
nowhere does not follow. Except this there is one more minus in offered
explanation:
 if shadow
photon acts upon real moreover thereby that given influence is
reflected on results of
experiment -- exactly on interferentional picture moreover direct
image -- then must be
the
equations, describing this interaction.
Furthermore  and suggestion about
existance of such photons can be incorrect. Really, the fact of
that no sensors could
not fix presence of shadow photon as well as a straight line
dependency from it
received interferentional picture, results in conclusion about
a zero energy of its and, as
a result, zero stress-energy tensor.
All this seamingly speaks about impossibility like to situations.
But for full understanding
of the presenteded position let us postpone aside photons. After
all in begining the problem
stood in understanding of interferentional natures in general
quantum particles.
Simply Deutsch considered the case with photons.
Since the interference of quantum particles runs equally then
conclusion from \cite{1}
about parallel universes can generalise, for instance, on the
spinor fields. But
again for this we need the equations, describing a shadow spinor
fields, which, as was
spoken above, in view of zero energy have and zero stress-energy tensor.
But this now does not difficult problem! Because there are so-called
the ghost spinors!
Thereby, we get at first, a physical interpretation  of the ghost
spinors -- this is  a
corresponding spinor fields in the parallel universes, and at
secondally, a physical motivation
of shadow particles -- the equations its describing are found.
Moreover, we now
have got the possibility to do wholly legal transition to statement about
existance of parallel universes.
 Really,
since all bodies are consisted by atoms, but atoms are consisted
by electrons, neutrons and
protons, which are described by the Dirac's equation, then a
presence in space of ghosts for
these particles draws a presence of ghosts and for each body,
i.e. we get the set of parallel
worlds.

\section*{Conclusion.}
In this article the ghost spinors were found, and also its phisical
interpretation which takes for basis a parallel worlds is done.
But moreover
we have to say that hereafter the matter and substence are not a synonyms.
So-called ghosts as once are a matter and are not a substence,
i.e. there are a
flows of particles which not have a characters of the last.

Come back to the
question about photons we consider that shadow photons in change from real
photons are can not discribed by the Makswell's equations. This is because
from the zero stress-energy tensor following a absence of electric-magnetic
field in our spacetime.
But here contradictions are absented! After all the real photon is nor
than other as a
carrier of a certain energy, but energy a shadow photon is a zero.
As once in this fact
it consists
the difference between photons and particles with half-whole spin.
May be the problem is solved by the finding for photons the field
equations with
greater degrees of freedom than  the Makswell's equations.

%%%%%%%%%%%%%%%%%%%%%%%%%%%%%%%%%%%%%%%%%%

\end{document}